\documentclass[reprint, superscriptaddress, amsmath,amssymb,prb,twocolumn]{revtex4-2}
\usepackage[utf8]{inputenc}
\usepackage{epsfig}
\usepackage{graphicx}
\usepackage{adjustbox}
\usepackage{indentfirst}
\usepackage{multirow}
\usepackage{latexsym}
\usepackage{color, soul}
\usepackage{ragged2e}
\usepackage{upgreek}
 \usepackage{amsmath}
\usepackage{amsmath}
\usepackage{amssymb}
\usepackage{mathtools}

\usepackage{hyperref}
\providecommand{\url}[1]{\texttt{#1}}
\usepackage{dcolumn}
\usepackage{graphicx, xcolor}
\usepackage[normalem]{ulem}
\usepackage{sansmath}
\usepackage{float}

	\hypersetup{
	   	bookmarks		= false,		
		unicode			= false,	
		pdfborder			= {0 0 0}	
		pdftoolbar			= false,		
		pdfmenubar		= true,		
		pdffitwindow		= false,     	
		pdfstartview		= {FitH},    	
		pdfnewwindow	= true,      	
		colorlinks			= true,		
		linkcolor			= blue,		
		citecolor			= blue,		
		filecolor			= blue,		
		urlcolor			= blue,		
		linktocpage		= true,		
		pdftitle				= {Using In-Plane Anisotropy to Engineer Janus Monolayers of Rhenium Dichalcogenides},
		pdfauthor			= {M. Mucha-Kruczynski},
		pdfsubject			= {},
		pdfcreator			= {MikTeX},
		pdfproducer		= {Marcin Mucha-Kruczynski},
		pdfkeywords		= {},
		}

\newcounter{bla}

\usepackage[normalem]{ulem}
\def\gap{HSE gap}

\def\SSex{ReS$_{2-x}$Se$_x$}
\def\SSeOne{ReS$_{1.75}$Se$_{0.25}$}
\def\SSeTwo{ReS$_{1.5}$Se$_{0.5}$}
\def\SSeThree{ReS$_{1.25}$Se$_{0.75}$}
\def\SeSx{ReSe$_{2-x}$S$_x$}
\def\SeSOne{ReSe$_{1.75}$S$_{0.25}$}
\def\SeSTwo{ReSe$_{1.5}$S$_{0.5}$}
\def\SeSThree{ReSe$_{1.25}$S$_{0.75}$}

\newcommand\SOne{1}
\newcommand\STwo{2}
\newcommand\SThree{3}
\newcommand\SFour{4}

\newcommand{\vect}[1]{\boldsymbol{#1}}	

\makeatletter
\DeclareMathOperator{\myhelper@Re}{Re} 
\renewcommand{\Re}{\myhelper@Re}
\DeclareMathOperator{\myhelper@Im}{Im} 
\renewcommand{\Im}{\myhelper@Im}
\makeatother

\begin{document}

\title{Using In-Plane Anisotropy to Engineer Janus Monolayers of Rhenium Dichalcogenides}
\author{Nourdine Zibouche}
\email{n.zibouche@bath.ac.uk}
\affiliation{Department of Chemistry, University of Bath, Bath BA2 7AY, United Kingdom\\}
\author{Surani M.~Gunasekera}
\affiliation{Department of Physics, University of Bath, Bath BA2 7AY, United Kingdom\\}
\affiliation{Centre for Nanoscience and Nanotechnology, University of Bath, Bath BA2 7AY, United Kingdom\\}
\author{Daniel Wolverson}
\affiliation{Department of Physics, University of Bath, Bath BA2 7AY, United Kingdom\\}
\affiliation{Centre for Nanoscience and Nanotechnology, University of Bath, Bath BA2 7AY, United Kingdom\\}
\author{Marcin Mucha-Kruczynski}
\email{m.mucha-kruczynski@bath.ac.uk}
\affiliation{Department of Physics, University of Bath, Bath BA2 7AY, United Kingdom\\}
\affiliation{Centre for Nanoscience and Nanotechnology, University of Bath, Bath BA2 7AY, United Kingdom\\}

\begin{abstract}
The new class of Janus two-dimensional (2D) transition metal dichalcogenides with two different interfaces is currently gaining increasing attention due to the possibility to access properties different from the typical 2D materials. Here, we show that in-plane anisotropy of a 2D atomic crystal, like ReS$_{2}$ or ReSe$_{2}$, allows formation of a large number of inequivalent Janus monolayers. We use first-principles calculations to investigate the structural stability of 29 distinct ReX$_{2-x}$Y$_{x}$ ($\mathrm{X,Y \in  \{S,Se\}}$) structures, which can be obtained by selective exchange of exposed chalcogens in a ReX$_{2}$ monolayer. We also examine the electronic properties and work function of the most stable Janus monolayers and show that the large number of inequivalent structures provides a way to engineer spin-orbit splitting of the electronic bands. We find that the breaking of inversion symmetry leads to sizable spin splittings and spontaneous dipole moments that are larger than those in other Janus dichalcogenides. Moreover, our calculations suggest that the work function of the Janus monolayers can be tuned by varying the content of the substituting chalcogen. Our work demonstrates that in-plane anisotropy provides additional flexibility in sub-layer engineering of 2D atomic crystals.
\end{abstract}

\maketitle

\newpage

\section{Introduction}


Symmetry is a powerful unifying principle in science and the study of symmetry breaking is one of the central questions in physics. In Janus materials, broken symmetry is made explicit by the formation of two interfaces with different chemical compositions and, hence possibly, dissimilar properties \cite{walther_chemrev_2013}. While such a concept was initially realised at the micrometer scale with spherical glass beads \cite{casagrande_epl_1989}, more recently, atomically thin Janus materials became reality by engineering of two-dimensional (2D) atomic crystals like graphene and transition metal dichalcogenides (TMDs) \cite{lu_natnano_2017, zhang_acsnano_2017, li_small_2018, Sant_2020, yagmurcukardes_apr_2020, zhang_jmca_2020}. The latter are built of sandwich-like layers in which an atomic plane of transition metals is embedded between two planes of chalcogen atoms, and it has been shown that it is possible to produce Janus TMD layers in which these outer planes are made of different chalcogens. In their pioneering work, Lu {\it{et al.}} have successfully synthesized Janus MoSSe with a spontaneous out-of-plane dipole by selenization of MoS$_{2}$ \cite{lu_natnano_2017} while Zhang {\it{et al.}} have fabricated a monolayer Janus MoSeS by sulfurization of MoSe$_2$ \cite{zhang_acsnano_2017}. More recently, Sant {\it{et al.}} have also demonstrated a monolayer Janus PtSeS by sulfurization of PtSe$_2$ \cite{Sant_2020}. Janus 2D materials present an interesting contrast to the idea of stacking 2D materials on top of each other into van der Waals heterostructures; \cite{geim_nature_2013} here instead, one engineers a single monolayer crystal. Janus TMDs in particular have several interesting properties including a strong Rashba spin splitting, second-harmonic response, piezoelectricity and good catalytic performance \cite{li_small_2018, yagmurcukardes_apr_2020, zhang_jmca_2020, yuan_pccp_2017}.

\begin{figure}[!t]
\begin{center}
\includegraphics[width=1.00\columnwidth]{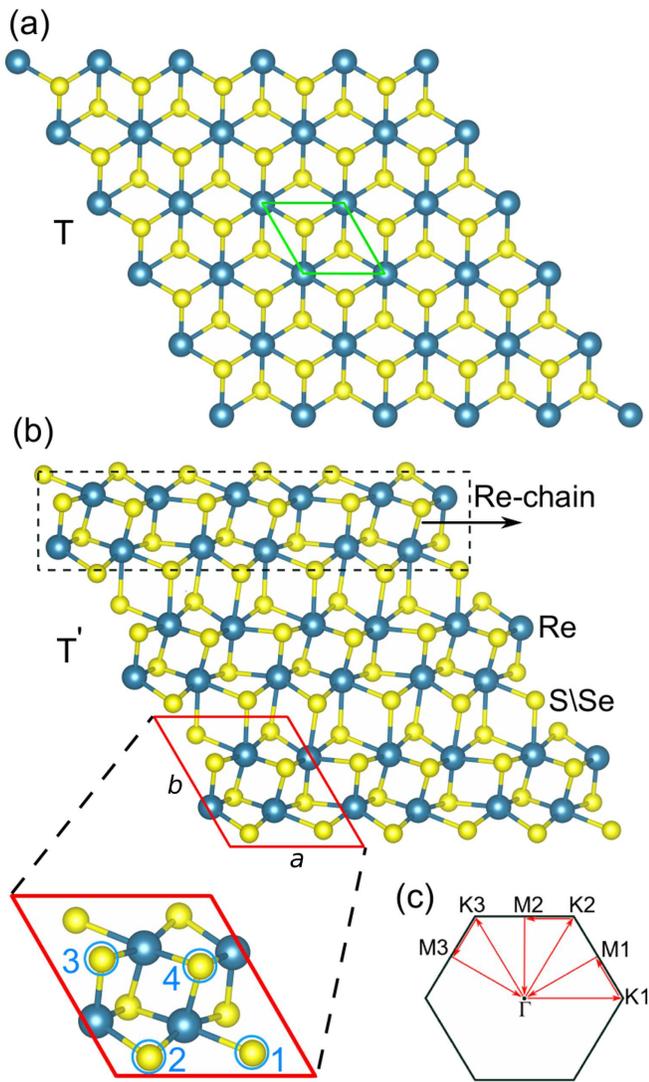}
\caption{Top view of (a) ideal trigonal structure (T) of monolayer transition metal dichalcogenides; (b) distorted trigonal phase (T$^\prime$) of ReX$_2$, X = S, Se, with green (yellow) balls indicating the positions of rhenium (chalcogens). The green and red rhombi indicate the unit cells of the two structures, with $a$ and $b$ marking the lattice constants for the T$^\prime$ structure. The inset in (b) highlights the four inequivalent chalcogen sites on the same chalcogen sub-layer, which are numbered from 1 to 4 and referred to as \SOne{}, \STwo{}, \SThree{}, and \SFour{}, respectively. The dashed rectangle marks the rhenium chains in the T$^\prime$ phase along the horizontal direction. (c) Brillouin zone of monolayer ReX$_2$ highlighting the full $k$-path used to plot the band structures in Fig.~\ref{fig:phonons} and \ref{fig:bands}.}
\label{fig:structures}
\end{center}
\end{figure}

In most TMDs, all of the chalcogen sites in the same plane are equivalent due to the combination of $C_{3}$ and translational symmetries present in the most common trigonal prismatic and octahedral phases \cite{manzelli_nrm_2017}. We use the example of rhenium-based TMDs to show that 2D materials with lower symmetry provide an additional flexibility in the design of Janus materials, as inequivalence of atomic sites translates into potential chemical and physical site selectivity. In the case of ReS$_{2}$ and ReSe$_{2}$, the ideal octahedral (T) arrangement of chalcogens around the transition metal sites, Fig.~\ref{fig:structures}(a), is distorted into characteristic rhenium chains \cite{lamfers_jac_1996, kertesz_jacs_1984}, as shown in Fig.~\ref{fig:structures}(b). In the new structure (T$^\prime$), breaking of the $C_{3}$ symmetry leaves inversion as the only non-trivial symmetry. We use density functional theory calculations to assess the energetic stability and electronic structure of ReS$_{2-x}$Se$_{x}$ and ReSe$_{2-x}$S$_{x}$ monolayer crystals obtained by exchanging chalcogen species in inequivalent sites in one of the chalcogen planes. We show that in rhenium TMDs this allows the formation of a large number of different Janus monolayers. We identify the most stable structures, discuss their dynamical stability and investigate their work function and electronic properties, including the spin-orbit splitting and out-of-plane dipole moment arising due to broken inversion symmetry in the Janus layers. Notably, in ReSSe we find the dipole moment to be about four times larger than that reported for other Janus TMDs \cite{riis-jensen_jpcc_2018, Xia_2018}.

Note that several different distortion-driven unit cells, some of which are only metastable, have been observed in TMDs \cite{manzelli_nrm_2017, qian_science_2014, sokolikova_csr_2020}. All of the distorted unit cells share the two features relevant here: (i) the $C_{3}$ symmetry of the ideal T phase is broken while (ii) the inversion symmetry is preserved. Hence, our conclusions regarding the potential richness of Janus monolayers of ReX$_{2-x}$Y$_{x}$ can be generalised to these other distorted structures.

\section{Methods}\label{sec.results}

Calculations are carried out using density functional theory (DFT) \cite{kohn_physrev_1965} as implemented in the Quantum Espresso simulation package \cite{gianozzi_jpcm_2009, gianozzi_jpcm_2017}. The Kohn-Sham wavefunctions and energies are calculated using the local density approximation (LDA) in the Perdew-Zunger parametrization \cite{pz_1981}, using a plane-wave basis with energy and charge density cutoffs of 60 and 360 Ry, respectively. Projector augmented-wave pseudopotentials \cite{PAW1994} are used to describe the core-valence interactions. The structural relaxation is performed until the force on each atom is smaller than 0.005 eV/\AA. The Brillouin Zone integration is sampled following the Monkhorst-Pack scheme \cite{monkhorst_prb_1976}, using 9$\times$9$\times$1 and 13$\times$13$\times$1 $k$-point grids for the ionic optimization and the electronic structure analysis, respectively. To avoid spurious interactions between periodically repeated slabs, the vacuum region is set to 25 \AA, and dipole corrections were included to eliminate artificial electrostatic effects. When calculating the formation energies, we have checked that for an enlarged unit cell (for example, with the lattice constant doubled from $a$ to $2a$), the conclusions regarding the energetic stability follow the same trend.

For more accurate band gaps, we have employed the HSE06 hybrid functional \cite{HSE06_2003, HSE06_2006}, instead of the more computationally expensive GW methods \cite{Hedin_1965, SLouie_2003,SGW2020}, starting from the exchange-correlation energy expressed by LDA, including spin-orbit coupling. 

The phonon band structures were calculated using density functional perturbation theory (DFPT) \cite{baroni_rmp_2001} in the Quantum Espresso code. The dynamic matrix was estimated on a $4\times4\times1$ $q$-point mesh in the Brillouin zone; such sampling is found to be sufficient for obtaining well-converged results. The dynamical matrices at arbitrary wave vectors were obtained using Fourier Transform-based interpolation to obtain phonon dispersion along the same wave vector path as the electronic band structures. 

The crystal orbital Hamilton populations (COHP) \cite{dronskowski_jpc_1993} presented in the Supplemental Material \cite{SM} were computed using the LOBSTER code \cite{maintz_jcc_2013, maintz_jcc_2016}. We employed the local basis function as given by Bunge \cite{bunge_adnd_1993} for the projection of the 3$s$ and 3$p$ orbitals for S, 4$s$ and 4$p$ for Se, and 5$s$, 6$s$, 5$p$, and 5$d$ for Re atoms. For the charge density analysis, we performed integrations of the electronic charge of the atoms following Bader charge method implemented in Henkelman's code \cite{henkelman_cms_2006, sanville_jcc_2007, tang_jpcm_2009, yu_jcp_2011}.

\section{Results and Discussion}
    \subsection{Energetic stability and structural properties}

We start our discussion with a closer look at a unit cell of monolayer ReX$_2$, $\mathrm{X=S,Se}$, as shown in the inset of Fig.~\ref{fig:structures}(b). Due to a Jahn-Teller distortion, the lattice vectors of the distorted T$^\prime$ structure are approximately twice as large as in the ideal T phase and the unit cell contains 12, rather than 3, atoms -- 4 rheniums and 8 chalcogens \cite{lamfers_jac_1996, kertesz_jacs_1984}. As inversion is the only point group symmetry element still present, all four of the chalcogen sites on the same side of the transition metals (either above or below) are inequivalent. We highlight one set of such sites in the inset of Fig.~\ref{fig:structures}(b) and number them as 1, 2, 3, and 4. The low symmetry of the unit cell of monolayer rhenium dichalcogenides results in a Brillouin zone which is a distorted regular hexagon, as shown in Fig.~\ref{fig:structures}(c), where we have labeled the inequivalent corners and centres of sides and marked a path in the momentum space used to plot the band structures later on.

In the bulk, the inequivalence of chalcogen sites has been observed explicitly by studying the Raman spectra of ReSe$_{2-x}$S$_{x}$ alloys containing low levels of sulfur \cite{hart_npj2d_2017} and investigating ReSe$_{2}$ using tunnelling microscopy \cite{hong_acsnano_2018}. In the former study, four different Raman bands arising from the local vibrational modes of sulfur atoms were observed, corresponding to substitutions occupying each of the sites. Interestingly, because in ReX$_2$ the valence band maximum is predominantly built from in-plane rhenium $d$-orbitals \cite{kertesz_jacs_1984, choi_acsnano_2020}, a unique coupling develops between the crystal thickness and in-plane anisotropy of the electronic dispersion: the anisotropy increases as the crystal is thinned down and reaches maximum in the monolayer \cite{hart_prb_2021}. This suggests that rhenium TMD monolayers are an ideal platform to explore how in-plane anisotropy of 2D crystals can be used to engineer novel atomically thin Janus materials.  

Recently, Janus TMD monolayers have been produced by exchange of chalcogens in the layer above the transition metals either by sulfurization of a selenide \cite{zhang_acsnano_2017} or selenization of a sulfide \cite{lu_natnano_2017}. In particular, Zhang {\it{et al.}} demonstrated a fine degree of control over the sulphurization of MoSe$_2$ as a function of the temperature of the process \cite{zhang_acsnano_2017}. An even greater degree of control was achieved by Lin {\it{et al.}} who used pulsed laser deposition \cite{lin_acsnano_2020} and tuned kinetic energy of Se plasma to determine conditions for the formation of high-quality Janus WSSe monolayers at temperatures below 300 $^{\circ}$C.

In hexagonal TMDs, such processes cannot distinguish between various chalcogen sites as they are related by the $C_{3}$ rotational symmetry and lattice translation. However, as in ReX$_{2}$ the chalcogen sites 1-4 are inequivalent, it could be possible to devise a process which will exchange the chalcogens selectively rather than en masse. In the rest of the paper, we use the notation ReX$_{2-x}$Y$_x$($i, ..., j$), with concentration $x\in(0.25,0.5,0.75,1)$, to denote a monolayer of ReX$_2$ in which 4$x$ of the X atoms in positions $i, ..., j\in(1,2,3,4)$ as marked in the inset of Fig.~\ref{fig:structures}(b) have been substituted with Y chalcogens. In this notation, $\tfrac{x}{2}$ is the fraction of all chalcogen sites that have been exchanged and, for example, ReSe$_{1.5}$S$_{0.5}$(1,3) denotes a ReSe$_2$ monolayer in which selenium atoms in positions 1 and 3 have been replaced by sulphur. For Janus layers in which all of the top four chalcogens have been substituted, the indices and substitution positions can be omitted without loss of clarity. Moreover, since in our calculations we model free-standing monolayers, the two ``full Janus'' structures, ReSeS and ReSSe, are structurally and electronically equivalent. In total, we obtain 29 distinct Janus structures and we present the formation energies for all of them as well as those of pristine ReS$_{2}$ and ReSe$_{2}$ monolayers in Fig.~\ref{fig:energies} (in the Supplemental Material \cite{SM}, we also present the data from Fig.~\ref{fig:energies} in the form of the binary convex hull, Fig.~S1, in which the formation energies of the Janus monolayers are directly compared to those of the pristine ReS2 and ReSe2 monolayers).

\begin{figure}[!t]
\begin{center}
\includegraphics[width=1.0\columnwidth]{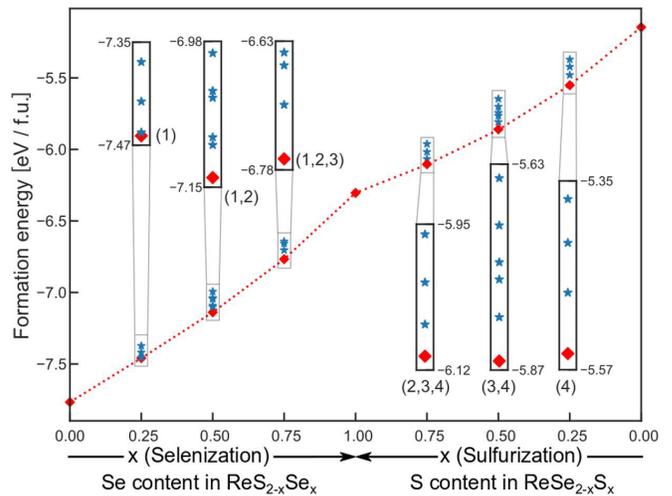}
\caption{Formation energies of 29 \SSex{} and \SeSx{} Janus monolayers of rhenium dichalcogenides considered in this work as well as the pristine ReS$_{2}$ and ReSe$_{2}$. In the left half of the Figure, number of Se atoms in the top chalcogen layer increases from left ($x=0$) to right ($x=1$). In the right half of the Figure, number of S atoms in the top chalcogen layer increases from right ($x=0$) to left ($x=1$). The insets zoom in on the formation energies of layers with the same chemical formula. The red diamond denotes the most stable structure for a given S/Se concentration and the numbers in brackets identify the corresponding atomic configuration.}
\label{fig:energies}
\end{center}
\end{figure}

To evaluate the energetic stability of our structures, we have calculated the formation energies $E_{\mathrm{f}}$ per formula unit (f.~u.),
\begin{align}\begin{split} \label{eq:nrg}
E_{\mathrm{f}}(\mathrm{ReX}_{2-x}\mathrm{Y}_x) = &\, E_{\mathrm{Tot}}(\mathrm{ReX}_{2-x}\mathrm{Y}_x) - E_{\mathrm{Tot}}(\mathrm{ReX}_2)  \\
&- (2-x)E_{\mathrm{Tot}}(\mathrm{X}) + xE_{\mathrm{Tot}}(\mathrm{Y}), 
\end{split}\end{align}
where $E_{\mathrm{Tot}}$ is the total energy of each compound and $\mathrm{(X,Y)\in(S,Se)}$ are the chalcogen species. For the pure elements S and Se, we have taken as reference formation energies of their most stable crystalline phases at ambient conditions (monoclinic for Se and orthorhombic for S). The results in Fig.~\ref{fig:energies} show that the pristine ReS$_{2}$ and ReSe$_{2}$ monolayers have, respectively, the lowest (-7.77~eV per f.~u.) and the highest (-5.15~eV per f.~u.) formation energies among all the investigated structures. The formation energies of all the Janus layers fit between these two values, with energetic stability decreasing (and the value of $E_{\mathrm{f}}$ increasing) linearly with the increasing of the selenium content. This suggests that it is energetically more convient to engineer the Janus monolayers by sulfurization of ReSe$_{2}$ rather than selenization of ReS$_{2}$.  

The formation energy of a full Janus layer ReSSe is -6.30~eV, which is close to the mean value (-6.46~eV) of the formation energies of the pristine monolayers. We have checked that, for ReSSe, our calculations predict the same formation energy, irrespectively of whether we consider a selenized ReS$_{2}$ or a sulphurized ReSe$_{2}$. All other structures with the same chalcogen concentrations have different formation energies because, as suggested by the in-plane anisotropy of rhenium dichalcogenides, exchange of chalcogens on each of the four inequivalent sites corresponds to dissimilar energy costs. The formation energies of the \SSeOne{}($i$) structures, $x=0.25$ in the left half of Fig.~\ref{fig:energies}, show that \SSeOne{}(1) is the most stable ($E_{\mathrm{f}}=-7.46$~eV per unit cell), followed by \SSeOne{}(2), \SSeOne{}(3) and \SSeOne{}(4) with formation energies higher by 4 meV, 40 meV, and 86 meV, respectively. Similar site ordering, $1\rightarrow2\rightarrow3\rightarrow4$, is seen for a step-by-step selenization of ReS$_{2}$ into ReSSe as \SSeOne{}(1), \SSeTwo{}(1,2) and \SSeThree{}(1,2,3) are the most stable structures for $x=0.25$, $x=0.5$ and $x=0.75$, respectively. To note, the very small energy difference between \SSeOne{}(1) and \SSeOne{}(2) means that both configurations are likely to form for the $x=0.25$ of selenide concentration. Notably, as seen from the location of sites 1 and 2 in Fig.~\ref{fig:structures}(b), exchange of sulfur to selenide is preferential at the chalcogen sites between the rhenium chains. This is in agreement with the observations of preferential Se locations in ReS$_{2-x}$Se$_{x}$ alloy \cite{wen_nanoscale_2017}. Overall, we find that for a Se atom substituting sulfur, the distances to its three nearest rhenium atoms increase by as much as 5.5\%. 

\begin{figure}[!t]
\begin{center}
\includegraphics[width=1.0\columnwidth]{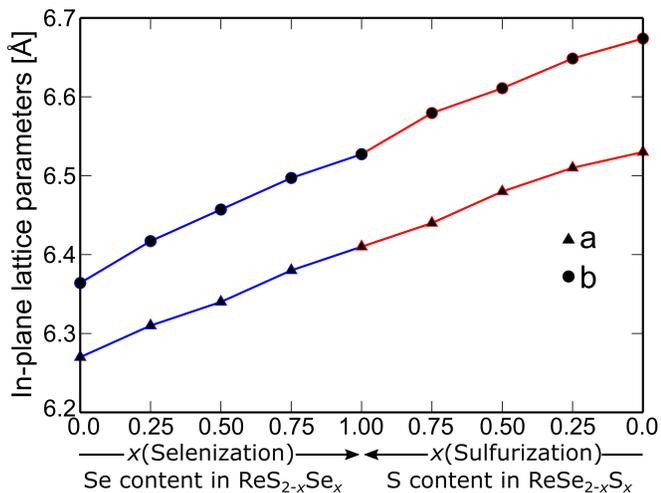}
\caption{In-plane lattice parameters $a$ (circles) and $b$ (triangles) of \SSex{} and \SeSx{} monolayers as a function of the concentration $x$. See Fig.~\ref{fig:structures}(b) for the schematic drawing of the unit cell.}
\label{fig:lattice}
\end{center}
\end{figure}

\begin{figure*}[bt]
\begin{center}
\includegraphics[width=1.0\textwidth]{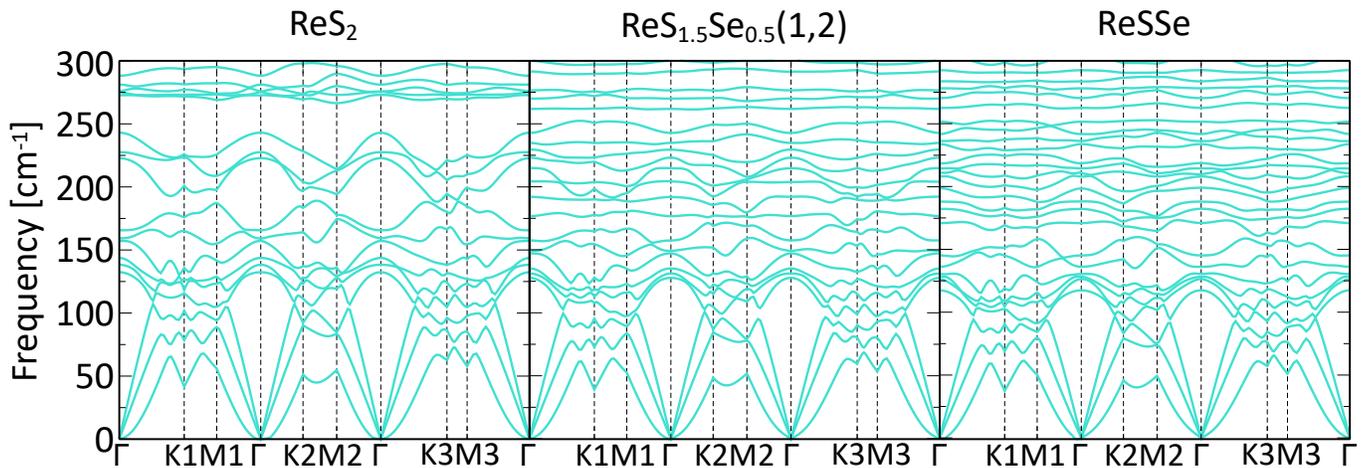}
\caption{Phonon spectra of ReS$_{2}$ (left), ReS$_{1.5}$Se$_{0.5}$(1,2) (centre) and ReSSe (right) along wave vector path as indicated in Fig.~\ref{fig:structures}.}
\label{fig:phonons}
\end{center}
\end{figure*}

The energy ordering of the chalcogen sites preferred during sulfurization of a ReSe$_{2}$ monolayer, right half of Fig.~\ref{fig:energies}, is reversed with respect to that discussed for ReS$_{2-x}$Se$_{x}$: the most stable configurations are \SeSOne{}(4), \SeSTwo{}(3,4), and \SeSThree{}(2,3,4) for $x=0.25$, $x=0.5$ and $x=0.75$, respectively. Similarly, for the \SeSOne{} structure, the energetic stability of the sites follows the sequence \SFour{}, \SThree{}, \STwo{}, and \SOne{} from the most to the least stable -- inversed as compared to that of the \SSeOne{} monolayer. Together with the earlier results for selenization of ReS$_{2}$, this means that the energetics of chalcogen substitution at any of the inequivalent sites can be considered effectively independent of what chalcogen species occupy the other sites. Moreover, such energetic ordering of the sites implies that sulfur prefers to substitute selenide at atomic positions located within the rhenium chains rather than between them. This can be explained by the difference in the size of the chalcogen species and the distorted structure of the T$^\prime$ phase. Due to the distortion, as compared to the T phase, the bond lengths between rhenium atoms and the chalcogen atoms between the chains (sites 1 and 2) are stretched while the bond lengths between rheniums and the chalcogens within the chain (sites 3 and 4) are shortened. Therefore, it is energetically easier to accommodate bigger chalcogen species (Se) between the chains as there is more space available as compared to within the chains. At the same time, the smaller chalcogen species (S) can fit within the rhenium chains with less effort against the local strain. We find that when one sulfur replaces a selenide atom, the bond length is shortened by 5.3\% on  site 1, compared to a value of 4.8\% on the other sites. To note, the energetic order of chalcogen vacancies in ReX$_{2}$ follows the sequence \SFour{}, \SThree{}, \STwo{}, \SOne{} \cite{horzum_prb_2014, zhu_iscience_2021}, the same as for exchange of Se to S. In both cases, the order is dictated by the preference for atomic sites corresponding to smaller volume in the lattice to minimize strain (either due to an empty site or substitution by an atom smaller than Se). The same mechanism reverses the site ordering for selenization of ReS$_{2}$ to accommodate larger Se atoms in place of smaller S.

We have investigated whether exchange of the chalcogen species leads to significant changes in the bond strength by computing the Crystal Orbital Hamilton Populations (COHP) \cite{dronskowski_jpc_1993} for the three shortest chalcogen-rhenium bonds for each of the chalcogen sites for ReS$_{2}$ and ReS$_{1.75}$Se$_{0.25}$(1). COHP partitions the band structure energy into orbital-pair interactions and can be interpreted as a ''bond-weighted'' density-of-states between a pair of atoms. It indicates the bonding character of states at given energy and its energy integral shows the contribution of a specific bond to the band energy and hence provides information about the bond strength. Our results presented in Fig.~S2 in the Supplemental Material \cite{SM} show little change in the COHP upon the exchange of one S to Se, indicating very similar bonds in both monolayers despite their significant lengthening in ReS$_{1.75}$Se$_{0.25}$(1) as compared to ReS$_{2}$. This suggests that interpretation of our results in terms of strain and chalcogen sites which can more readily accommodate the larger/smaller chalcogen species does capture the essential physics. 


The impact of the size of the chalcogen species on the structure of the Janus monolayer can also be seen by comparing their in-plane lattice constants, $a$ and $b$, shown in Fig.~\ref{fig:lattice}  for the pure monolayers and the most stable Janus crystals. Both vary nearly linearly between ReS$_{2}$ and ReSe$_{2}$ with chalcogen composition conforming to Vegard's law, which postulates the existence of a linear relationship between the crystal lattice constant of an alloy and the varying concentration of its constituents \cite{vegard_zphys_1921}. The lattice constants 
$a_{\mathrm{ReX}_{2-x}\mathrm{Y}_{x}}$, $b_{\mathrm{ReX}_{2-x}\mathrm{Y}_{x}}$ of a Janus monolayer ReX$_{2-x}$Y$_{x}$ can be determined as
\begin{align}
l_{\mathrm{ReX}_{2-x}\mathrm{Y}_{x}}=\left(1-\frac{x}{2}\right)l_{\mathrm{ReX}_{2}}+\frac{x}{2}l_{\mathrm{ReY}_{2}},
\end{align}
where $l=a,b$, $\mathrm{(X,Y)\in(S,Se)}$, {$a_{\mathrm{ReS}_{2}}=6.27$~\AA}, {$b_{\mathrm{ReS}_{2}}=6.36$~\AA}, {$a_{\mathrm{ReSe}_{2}}=6.53$~\AA} and {$b_{\mathrm{ReSe}_{2}}=6.67$~\AA}. For example, for the full Janus monolayer ReSSe, $a_{\mathrm{ReSSe}}=\tfrac{1}{2}(a_{\mathrm{ReS}_{2}}+a_{\mathrm{ReSe}_{2}})=6.40$~\AA, in agreement with our ab initio calculations.

Finally, apart from thermodynamic stability of the studied Janus monolayers, we also investigate their dynamic stability by studying their phonon spectra. In Fig.~\ref{fig:phonons}, we show the phonon band structures for three selected materials: pristine ReS$_{2}$, ReS$_{1.5}$Se$_{0.5}$(1,2) and ReSSe, calculated along the wave vector path as shown in Fig.~\ref{fig:structures}(c). We do not observe any appearance of imaginary energy solutions, indicating no instabilities or softening of the phonon modes. We suggest that this is because the low symmetry of the pristine ReX$_{2}$ compounds easily accommodates small distortions of bonds due to S$\leftrightarrow$Se substitutions. Spectra for all the structures are very similar in the range of small wavenumbers, $\tilde{\nu}\lesssim 150$ cm$^{-1}$. At higher wavenumbers, exchanging S to Se in ReS$_{2}$ lowers the energy of optical phonons which leads to vanishing of the gap present in ReS$_{2}$ around $\sim 250$ cm$^{-1}$ and flattening of their dispersion. This is due to the heavier mass of Se as compared to S \cite{hart_npj2d_2017}.


\begin{figure*}[!t]
\begin{center}
\includegraphics[width=\textwidth]{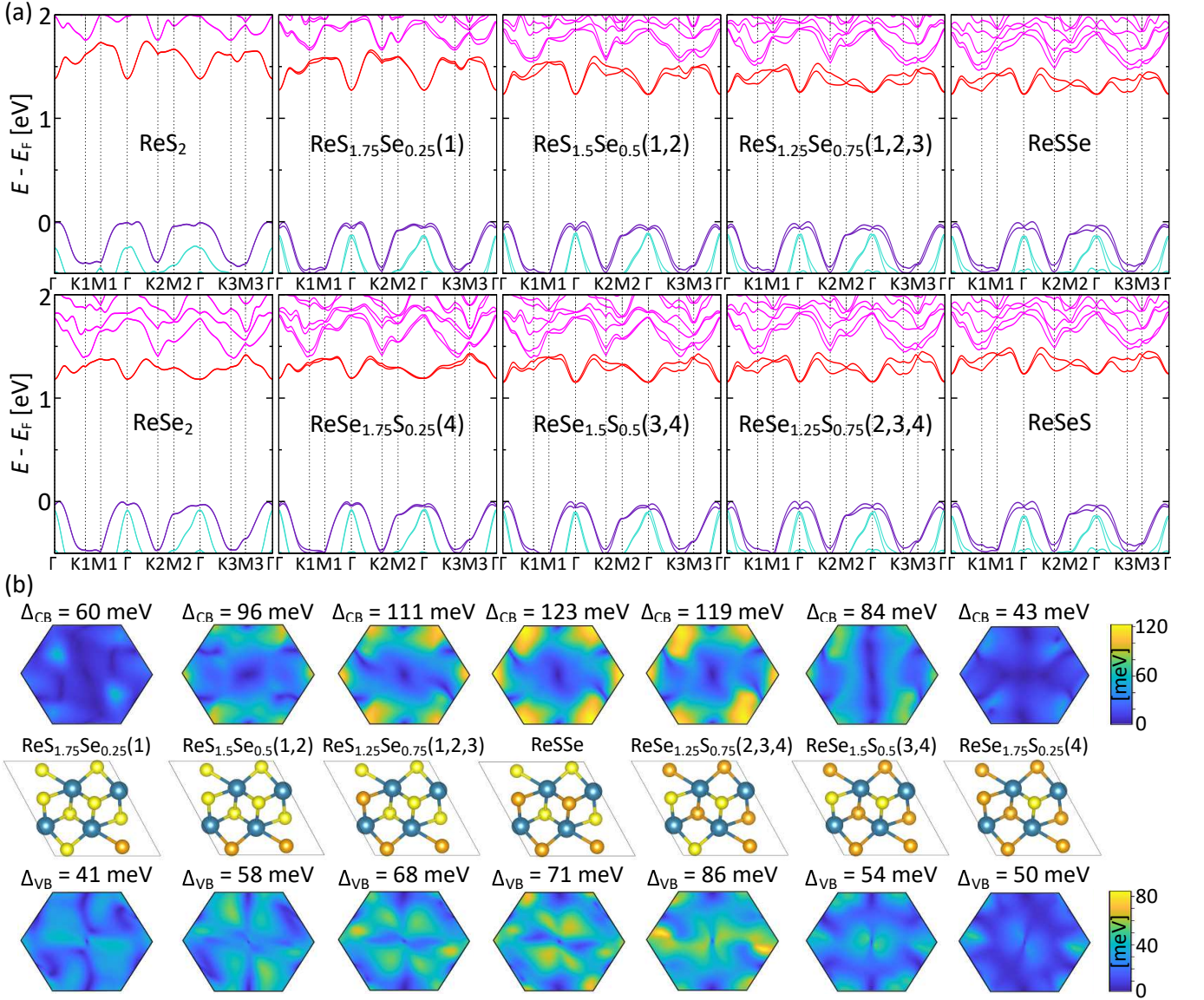}
\caption{(a) LDA band structures of \SSex{} (Top) and \SeSx{} (Bottom) monolayers calculated including spin-orbit coupling along wave vector path as indicated in Fig.~\ref{fig:structures}. The two uppermost valence sub-bands and the two lowest conduction sub-bands are shown in indigo and red, respectively. (b) Spin-orbit splitting of the lowest conduction band (top row) and the uppermost valence band (bottom) across the whole Brillouin zone; $\Delta_{\mathrm{CB}}$ and $\Delta_{\mathrm{VB}}$ denote the maximum splitting in each band for each structure. The top view of each structure is shown in the middle row.}
\label{fig:bands}
\end{center}
\end{figure*}

\subsection{Electronic properties} \label{sec:elec_prop}

Following our discussion of the stability of the various ReX$_{2-x}$Y$_{x}$ Janus monolayers, we discuss the electronic properties of only the most stable structures for each concentration. Figure~ \ref{fig:bands}(a) presents the band structures of these ReX$_{2-x}$Y$_{x}$ systems, from the pristine ReS$_{2}$ and ReSe$_{2}$ monolayers on the left (top and bottom row, respectively) to the full Janus structures, ReSSe and ReSeS, on the right. The momentum path through the Brillouin zone is chosen as indicated with the red arrows in Fig.~\ref{fig:structures}(c). As anticipated, the bands of ReSSe and ReSeS (right-most column) are identical. Due to the intrinsically present inversion symmetry, the bands of both the pristine ReS$_{2}$ and ReSe$_{2}$ monolayers (left-most column) are spin-degenerate. Also, within our computational framework, both are indirect band gap semiconductors: the conduction band minimum (CBM) is found at $\Gamma$ whereas the valence band maximum (VBM) is located slightly away from it. This is in agreement with other works using similar methodology \cite{hart_prb_2021, zhong_prb_2015} although a direct band gap has been predicted at the GW level \cite{echeverry_prb_2018} (interestingly, the latter also predicts direct band gaps in the bulk while experiments suggest this is not the case \cite{webb_prb_2017, hart_scirep_2017, eickholt_prb_2018, gunasekera_jem_2018}). All of the Janus monolayers inherit the CBM location as well as the shape of the valence band around $\Gamma$ from the pristine structures and remain indirect band gap semiconductors. However, exchange of any chalcogens only on one side of the transition metals breaks the inversion symmetry so that spectra of all Janus monolayers display band splitting due to spin-orbit coupling. While in general this splitting  increases as the concentration of the substituting chalcogen (and hence the extent of inversion symmetry breaking) increases, for a given momentum it does not simply grow linearly in magnitude from either of the pristine structures to the full Janus monolayer. In Fig.~\ref{fig:bands}(b), we show the momentum-resolved 2D maps of the spin-splitting of the non-degenerate lowest conduction and highest valence bands across the whole Brillouin zone and list the maximum conduction and valence band splittings, $\Delta_{\mathrm{CB}}$ and $\Delta_{\mathrm{VB}}$, respectively, for all the Janus structures. Notice that, in the valence band in particular, the regions of the Brillouin zone with the largest splitting change from one structure to the next and are not always found in the vicinity of high-symmetry directions in the Brillouin zone. Moreover, the largest possible valence band splitting $\Delta_{\mathrm{VB}}$ is produced in ReSe$_{1.25}$S$_{0.75}(2,3,4)$ rather than in the full Janus monolayer ReSSe. This suggests sub-layer atomic engineering of two-dimensional crystals as a potential strategy of tailoring the momentum-dependence of spin-orbit coupling.

While the band structures presented in Fig.~\ref{fig:bands} have been obtained in the LDA approximation, this method is well-known to provide underestimated band gaps. We performed additional calculations with HSE-type hybrid functional to produce more reliable band gap estimates. We list the LDA and HSE values for the most stable \SSex{} and \SeSx{} monolayers in the first two columns of Table~\ref{my-gaps}. Overall, the LDA band gaps are enlarged by about 50\% when HSE is used. Upon substitution of ReS$_{2}$ (ReS$_{2}$) with Se (S), the band gap decreases (increases), reaching the value of 1.86~eV for the full Janus monolayer. However, unlike the lattice constant, which follows Vegard's law, the the band gap does not show a linear change with concentration since the band gaps of \SSeThree{} and \SeSThree{} are sightly larger than or equal to those of \SSeTwo{} and \SeSTwo{}, respectively (See Table~\ref{my-gaps}). 

\begin{table}[!b]
\centering
\caption{LDA and HSE band gaps (in eV) as well as magntiude of the dipole moment, $|\vect{p}|$, (in Debye) of \SSex{} and \SeSx{} monolayers. The band gap values are calculated with spin-orbit coupling taken into account.}
\label{my-gaps}
\begin{tabular}{cccc}
 Structure   & \multicolumn{1}{c}{LDA~gap} & \multicolumn{1}{c}{\gap{}} & \multicolumn{1}{c}{$|\vect{p}|$} \\
\hline
ReS$_2$      &  1.39 & 2.06 &  0.00  \\
\SSeOne{}(1)    &  1.28 & 1.92 &  0.14  \\
\SSeTwo{}(1,2)    &  1.24 & 1.88 &  0.30  \\
\SSeThree{}(1,2,3)  &  1.26 & 1.89 &  0.50  \\
ReSSe / ReSeS  &  1.24 & 1.86 &  0.71  \\
\SeSThree{}(2,3,4)  &  1.16 & 1.76 &  0.56  \\
\SeSTwo{}(3,4)    &  1.16 & 1.76 &  0.39  \\
\SeSOne{}(4)    &  1.19 & 1.79 &  0.21  \\
ReSe$_2$     &  1.18 & 1.78 &  0.00  \\
\end{tabular}
\end{table}

For the pristine ReS$_{2}$ and ReSe$_{2}$, our LDA gap values are in a good agreement with the previously reported DFT calculations \cite{hart_prb_2021, zhong_prb_2015}. The more accurate values for the band gaps of the pristine single layers have been also calculated using the GW method, yielding 2.38~eV \cite{zhong_prb_2015, echeverry_prb_2018} and 2.09~eV \cite{zhong_prb_2015} or 2.05~eV \cite{echeverry_prb_2018} for ReS$_{2}$ and ReSe$_{2}$, respectively. Our HSE calculations underestimate these GW values by approximately 13\%.

\begin{figure}[!t]
\begin{center}
\includegraphics[width=1.0\columnwidth]{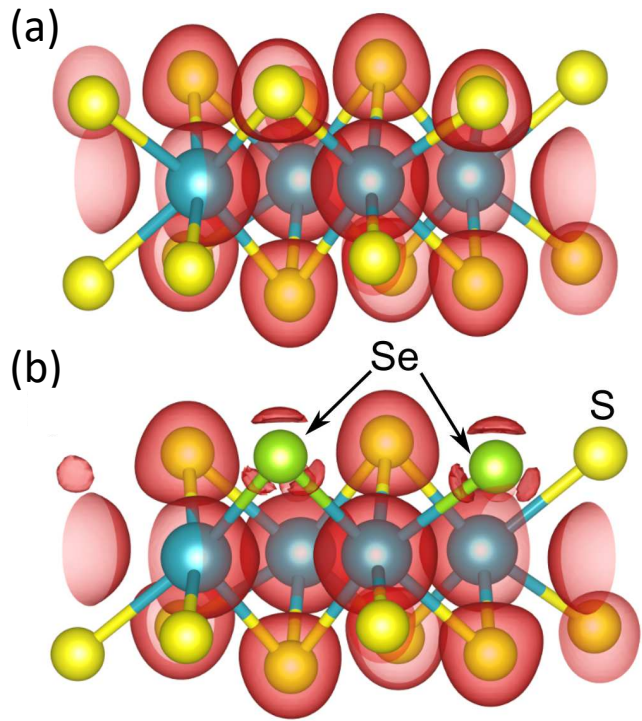}
\caption{Topology of the electron charge density distribution in (a) pristine ReS$_2$ and (b) \SSeTwo{} monolayers plotted with the same isovalue.}
\label{fig:charge}
\end{center}
\end{figure}

Another consequence of inversion-symmetry breaking is the formation of an out-of-plane dipole moment, $\vect{p}$, which arises due to the difference in electronegativities between the S and Se species \cite{lu_natnano_2017, ji_jpcc_2018, Xia_2018, riis-jensen_jpcc_2018} and points towards the chalcogen layer with a greater concentration of selenium. The magnitude of $\vect{p}$ increases with increasing concentration of the substituting chalcogen, $x$, reaching maximum, $|\vect{p}|=0.71\,\mathrm{D}=0.15\,e$\AA, where $e$ is the magnitude of electron charge, in the full Janus monolayer. We list the dipole moment for the most stable structures (as well as, for explicit comparison, for the pristine structures which have no dipole moment) in Table~\ref{my-gaps}. We compare the electron density distribution in the pristine ReS$_{2}$ ($|\vect{p}|=0$) and \SSeTwo{}(1,2) ($|\vect{p}|=0.3$ D) in Fig.~\ref{fig:charge}. Incorporating Se in place of S leads to a decrease of the electron density around the respective Se atomic sites and a transfer of charge, mainly to the rhenium atoms. The electron density around the S sites in the same chalcogen layer as the Se atoms is effectively unchanged. As shown by Bader charge analysis \cite{bader_charge, henkelman_cms_2006} presented in Table S1 of the Supplemental Material \cite{SM}, the charge redistribution due to S$\leftrightarrow$Se exchange at a given site is only weakly affected by the changes in the atomic species occupying chalcogen sites in the vicinity. We find that the dipole moment values of Janus ReSSe/ReSeS are about four times as large as the values of their MoSSe and WSSe analogues \cite{Xia_2018, riis-jensen_jpcc_2018}.


\subsection{Work function}

\begin{figure}[!t]
\begin{center}
\includegraphics[width=1.0\columnwidth]{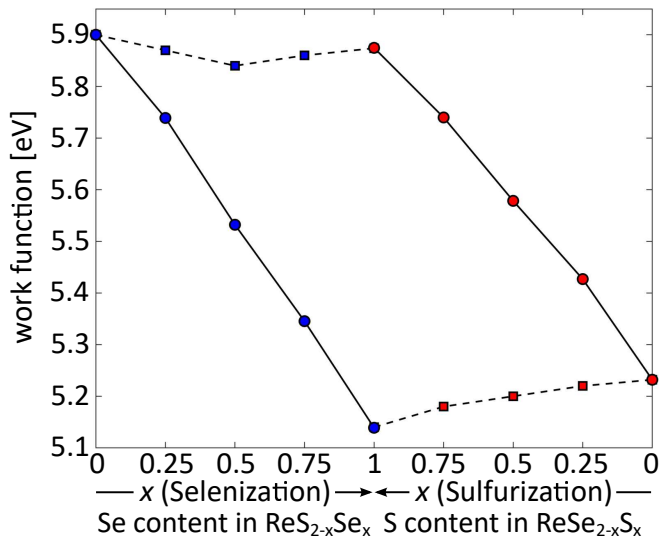}
\caption{Work function of \SSex{} and \SeSx{} monolayers as a function of the concentration of the substituted chalcogen $x$. The blue and red symbols represent the  \SSex{} and \SeSx{} systems, respectively. The circles denote the work function of the (top) side in which the chalcogens are being exchanged. The squares indicate the work function of the (bottom) side composed entirely of S for \SSex{} and Se for \SeSx{}.}
\label{fig:work_funk}
\end{center}
\end{figure}

The exchange of chalcogen atoms also modifies the work function of the material. In Fig.~\ref{fig:work_funk}, we show using the blue and red circles the work function, $\Phi$, taken here as the difference between the vacuum level and the valence band maximum, for all of the stable ReX$_{2-x}$Y$_{x}$ monolayers. We observe two nearly linear curves which show the work function for selenized ReS$_{2}$ (blue circles) and sulfurized ReSe$_{2}$ (red circles), respectively. For the former, the work function decreases linearly with the increasing concentration of Se, from 5.90 eV for ReS$_{2}$ to 5.14 eV for the full Janus monolayer ReSSe, close to the work function of the pristine ReSe$_2$ (5.24 eV). Inversely, the work function of sulfurized ReSe$_{2}$ increases linearly to the value of 5.87~eV for the full Janus monolayer ReSeS. The discontinuity in $\Phi$ in the centre of the graph is a consequence of the out-of-plane dipole moment $\vect{p}$. Recall that we assume here that exchange of the chalcogens takes place only in the layer above the transition metals which corresponds to the experimental situation of a monolayer crystal resting on a substrate with its top side exposed. As a result, the full Janus monolayer ReSSe obtained from ReS$_{2}$ and ReSeS obtained from ReSe$_{2}$ have dipole moments $\vect{p}$ oriented in opposite directions with respect to the surface which leads to the difference in work functions (our rigid choice of "up" and "down" breaks the inversion symmetry which otherwise links the two structures). Therefore, as pointed out before \cite{riis-jensen_jpcc_2018}, Janus TMDs provide interesting opportunities when considering band alignment in vertical or planar heterojunctions. For completeness, we also show in Fig.~\ref{fig:work_funk} the work functions computed for the bottom surfaces of all structures, shown using blue and red squares (and a black dashed line as a guide for the eye) for selenized ReS$_{2}$ and sulfurized ReSe$_{2}$, respectively. For a given structure, the difference between the work functions indicated with a square and a circle corresponds to the difference of electric potential energies of the two chalcogen layers.

\section{Conclusions}\label{sec.conclusions}
In summary, we have employed first-principles density functional simulations to study the stability as well as lattice and electronic properties of Janus monolayers of rhenium dichalcogenides. We have shown that the in-plane anisotropy of rhenium dichalcogenides leads to energetical inequivalence of Janus-like monolayers with the same sulfur and selenium content. While the selenium atoms prefer occupation of the interchain chalcogen sites, sulfur favours intrachain sites instead. In general, 2D alloys are thermodynamically preferred over their 2D Janus counterparts where S and Se atoms are separated from each other on either side of the (top or bottom) surfaces. However, recent experimental work demonstrates significant improvement of the understanding and degree of control of the chalcogen exchange process, aimed at preventing alloying and realizing Janus geometry \cite{lu_natnano_2017, zhang_acsnano_2017, Sant_2020, lin_acsnano_2020, qin_advmat_2022, jang_npgasiamater_2022}. This, coupled with the anisotropy-induced inequivalence of the chalcogen sites means that, at least in principle, in ReX$_{2}$ it could be possible to selectively exchange only a fraction of the chalcogens in a layer in contrast to all of them as is the case for other Janus TMDs. Such a partial chalcogen substitution provides a way to tune the work function of the material. Moreover, it enables a gradual breaking of inversion symmetry and hence engineering of spin-orbit splitting of the electronic bands in a 2D material. Finally, inversion symmetry breaking induces out-of-plane dipoles which in the full Janus monolayer are significantly larger than those reported for other Janus TMDs. Overall, our findings highlight the unusual properties of Janus rhenium dichalcogenides and, more generally, the relevance of in-plane anisotropy when engineering 2D materials.


\begin{acknowledgments}
This work was supported by the U.K. Engineering and Physical Sciences Research Council (EPSRC) through the Centre for Doctoral Training in Condensed Matter Physics,
Grant EP/L015544/1. We acknowledge use of the Balena High Performance Computing (HPC) Service at the University of Bath as well as the Isambard UK National Tier-2 HPC Service operated by GW4 and the UK Met Office and funded by EPSRC (Grant EP/P020224/1). For the purpose of open access, the authors have applied a Creative Commons Attribution (CC BY) licence to any Author Accepted Manuscript version arising. The data used in this study can be reproduced using the files available from the University of Bath data archive at \href{https://doi.org/10.15125/BATH-001136}{https://doi.org/10.15125/BATH-001136} \cite{data_archive}.
\end{acknowledgments}

\end{document}



\title{Supplemental Material: \\ Using In-Plane Anisotropy to Engineer Janus Monolayers of Rhenium Dichalcogenides}
\author{Nourdine Zibouche}
\email{n.zibouche@bath.ac.uk}
\affiliation{Department of Chemistry, University of Bath, Bath BA2 7AY, United Kingdom\\}
\author{Surani M.~Gunasekera}
\affiliation{Department of Physics, University of Bath, Bath BA2 7AY, United Kingdom\\}
\affiliation{Centre for Nanoscience and Nanotechnology, University of Bath, Bath BA2 7AY, United Kingdom\\}
\author{Daniel Wolverson}
\affiliation{Department of Physics, University of Bath, Bath BA2 7AY, United Kingdom\\}
\affiliation{Centre for Nanoscience and Nanotechnology, University of Bath, Bath BA2 7AY, United Kingdom\\}
\author{Marcin Mucha-Kruczynski}
\email{m.mucha-kruczynski@bath.ac.uk}
\affiliation{Department of Physics, University of Bath, Bath BA2 7AY, United Kingdom\\}
\affiliation{Centre for Nanoscience and Nanotechnology, University of Bath, Bath BA2 7AY, United Kingdom\\}


\maketitle

\newpage
\beginsupplement

%
%

\section{Binary convex hull}

\begin{figure}[t]
\begin{center}
\includegraphics[width=0.80\columnwidth]{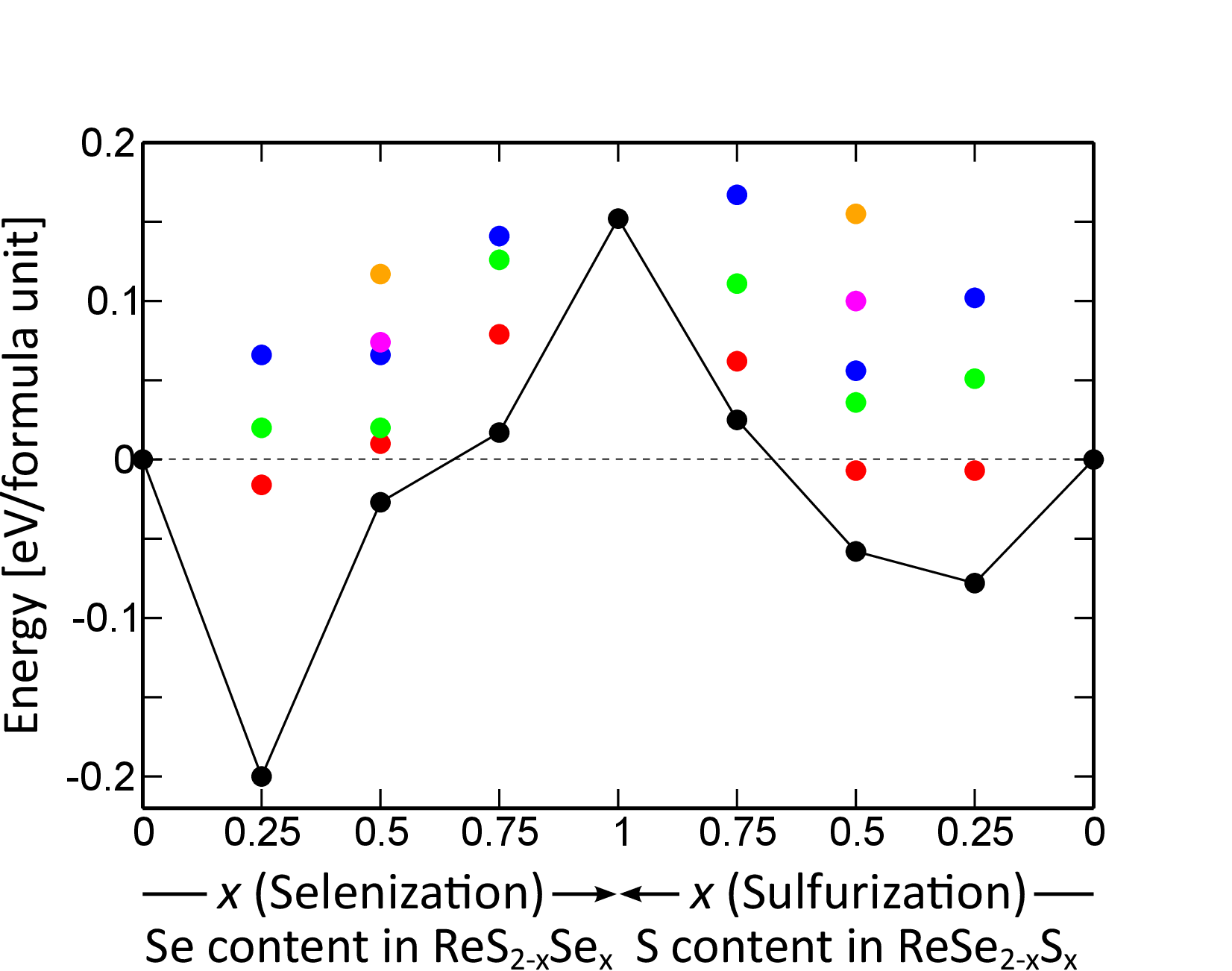}
\caption{Binary convex hull of ReX$_{2-x}$Y$_{x}$. The black dots denote the most stable structures for a given S/Se concentration. The other colours correspond to the other structures with their overall energetic ordering for a set concentration the same as described in the main text.}
\label{fig:convex_hull}
\end{center}
\end{figure}

In Fig.~\ref{fig:convex_hull}, we present data from Fig.~2 of the main text after subtracting a linear term so that the formation energies of both pure compounds are scaled to zero. This results in a binary convex hull which allows us to determine whether a particular Janus monolayer is energetically favourable to form as compared to the pristine materials -- this is the case if its formation energy lies below zero (dashed line in the figure). We denote with black dots the most stable compounds for a given S/Se concentration and use other colours for the remaining structures (the energy ordering of the structures for a set concentration is the same as described in the main text). It is clear that a small addition of S to ReSe$_{2}$ or Se to ReS$_{2}$ stabilizes the structure. Interestingly, the full Janus layer, ReSSe, lies the highest above zero which suggests that it is the least stable. However, this simply indicates that the ReSSe Janus monolayer would not form in a $1:1:1$ mixture of Re, S and Se but, rather, some other allotrope structure could result. Note that here, we focus on the stability and properties of materials prepared in a specific way starting from a single layer of a transition metal dichalcogenide. The Janus monolayers of transition metal dichalcogenides are not grown using any typical crystal growth method but rather obtained in a process which strips the top layer of chalcogens to allow their substitution with another \cite{lu_natnano_2017, zhang_acsnano_2017, Sant_2020}. The final material preserves the planar configuration and the process itself puts strict constraints on which lattice sites are available for chalcogen substitution (only on one side of the transition metals) so that Janus layers form even if they are only metastable. 

\section{Crystal orbital Hamilton populations analysis}

In order to investigate the changes in bonding occuring in Janus ReS$_{2-x}$Se$_{x}$ due to S$\leftrightarrow$Se exchange, we compute the energy-resolved (projected) Crystal Orbital Hamilton Populations (COHP) \cite{dronskowski_jpc_1993} for these materials,
\begin{align}
\mathrm{COHP}_{ij}(\epsilon) = H_{ij}\rho_{ij}(\epsilon),
\end{align}
where $H_{ij}$ is the Hamiltonian matrix element between Bloch states constructed from orbitals $i$ and $j$ and $\rho_{ij}(\epsilon)$ is the corresponding matrix element of the density of states matrix,
\begin{align}
\rho_{ij}(\epsilon)=\sum_{n}f_{n}c_{n,i}^{*}c_{n,j}\delta(\epsilon-\epsilon_{n}).
\end{align}
Above, $f_{n}$ and $\epsilon_{n}$ are the occupation factor and energy of band $n$, $c_{n,j}$ is the expansion coefficient of the wave function for the Bloch state composed from orbital $j$ and $\delta(\epsilon-\epsilon_{n})$ is the Dirac delta function in energy. COHP partitions the band structure energy into orbital-pair interactions and can be interpreted as a ''bond-weighted'' density-of-states between a pair of adjacent atoms. Whereas the electronic density of states shows where the electrons are but adds nothing about their bonding character, COHP indicates bonding, nonbonding and antibonding contributions to the band structure. For a bonding contribution, the corresponding Hamiltonian off-site element (and hence COHP) is negative, indicating lowering of the energy by forming a bond. Consequently, antibonding interactions are indicated by positive off-site Hamiltonian elements and COHP. Finally, while integrating the electronic density of states gives the number of electrons in the system, an energy integral of the COHP shows the contribution of a specific bond to the band energy so that integral of COHP provides information about the bond strength.

\begin{figure}[t]
\begin{center}
\includegraphics[width=1.00\columnwidth]{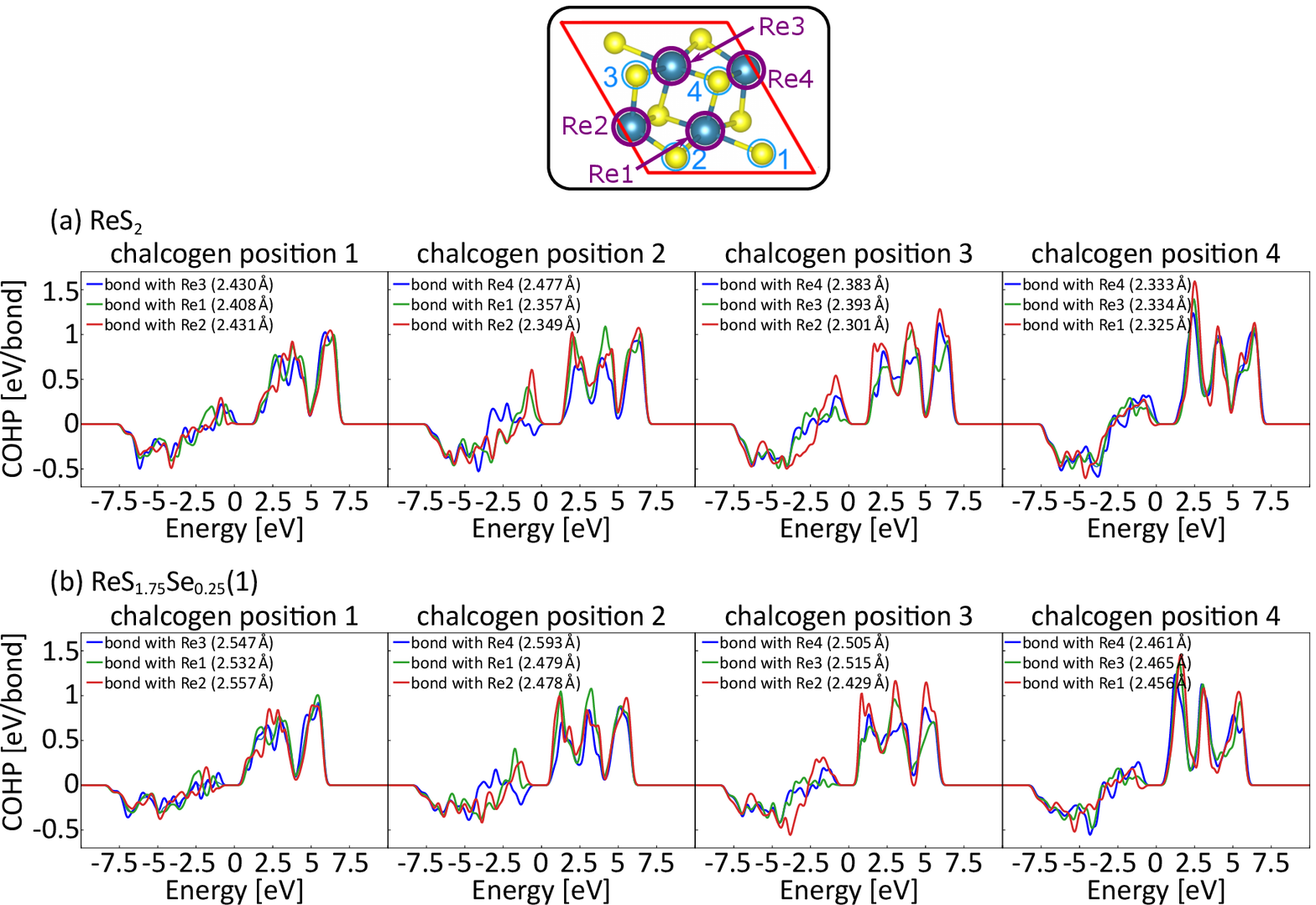}
\caption{Crystal Orbital Hamilton Populations (COHP) for chalcogen-transition metal bonds in (a) ReS$_{2}$ and (b) ReS$_{1.75}$Se$_{0.25}$. For each chalcogen site, 1-4, COHP for bonds with the three nearest Re sites (numbered like in the diagram on top of the figure) are shown.}
\label{fig:cohp}
\end{center}
\end{figure}

In Fig.~\ref{fig:cohp}, we show COHP calculated for the bonds between each of the four chalcogen sites and the nearest three of the four transition metals, both for ReS$_{2}$ (top row) as well as the most stable Janus monolayer with one of the S exchanged to Se, ReS$_{1.75}$Se$_{0.25}$(1) (bottom row). To identify each bond, in addition to the chalcogen sites numbered from 1 to 4 like in the main text, we number the rhenium sites from Re1 to Re4 as shown in the diagram on top of the figure (while two of the rhenium sites, X and Y, are related to the other two by inversion through the centre of the unit cell, for a given chalcogen site bonds with all transition metals are different). In the legend of each panel of Fig.~\ref{fig:cohp}, we also give the length of each bond. 

Overall, COHP plots show little change between a pure compound, ReS$_{2}$, and one of the Janus monolayers, ReS$_{1.75}$Se$_{0.25}$(1). Despite consistent lengthening of all chalcogen-rhenium bonds by at least 0.11 {\AA} after the exchange of S to Se at only one of the chalcogen sites, for both materials equivalent bonds show bonding/antibonding character at similar energies, although in ReS$_{1.75}$Se$_{0.25}$(1) both the bonding and antibonding features are consistently appearing at lower energies than in ReS$_{2}$. This confirms that the preference of the chalcogen sites for S$\leftrightarrow$Se exchange can be interpreted primarily in terms of the space available at a particular site versus the size of the chalcogen atom, as in the main text.

\section{Bader charge analysis}

\begin{center}
\begin{table}[b]
\caption{Bader charge (in the units of electron charge) for the pristine ReX$_{2}$ and energetically most stable rhenium Janus monolayers. Black, blue and red indicates, for each Janus monolayer, sites occupied by sulphur, rhenium and selenium atoms, respectively.} \label{bader_charge}
\scriptsize
\makebox[\textwidth]{\begin{tabular}{|c|c|c|c|c|c|c|c|c|c|} 
\hline
Site & ReS$_{2}$ & ReS$_{1.75}$Se$_{0.25}$(1) & ReS$_{1.5}$Se$_{0.5}$(1,2) & ReS$_{1.25}$Se$_{0.75}$(1,2,3) & ReSSe & ReSe$_{1.25}$S$_{0.75}$(2,3,4) & ReSe$_{1.5}$S$_{0.5}$(3,4) & ReSe$_{1.75}$Se$_{0.25}$(4) & ReSe$_{2}$ \\
\hline
1t & 6.544599 & \textcolor{red}{6.336945} & \textcolor{red}{6.345631} & \textcolor{red}{6.362025} & \textcolor{red}{6.371746} & \textcolor{red}{6.334817} & \textcolor{red}{6.348748} & \textcolor{red}{6.362706} & \textcolor{red}{6.374848} \\
2t & 6.474301 & 6.484028 & \textcolor{red}{6.266196} & \textcolor{red}{6.285994} & \textcolor{red}{6.305323} & 6.491471 & \textcolor{red}{6.271878} & \textcolor{red}{6.279836} & \textcolor{red}{6.294228} \\
3t & 6.453022 & 6.482132 & 6.483440 & \textcolor{red}{6.263021} & \textcolor{red}{6.278671} & 6.477553 & 6.489886 & \textcolor{red}{6.270704} & \textcolor{red}{6.286246} \\
4t & 6.429599 & 6.439648 & 6.465184 & 6.483046 & \textcolor{red}{6.264098} & 6.451559 & 6.465634 & 6.477053 & \textcolor{red}{6.260515} \\
Re1 & \textcolor{blue}{14.041145} & \textcolor{blue}{14.095648} & \textcolor{blue}{14.160576} & \textcolor{blue}{14.155069} & \textcolor{blue}{14.220043} & \textcolor{blue}{14.261307} & \textcolor{blue}{14.326605} & \textcolor{blue}{14.328686} & \textcolor{blue}{14.399420} \\
Re2 & \textcolor{blue}{14.057225} & \textcolor{blue}{14.083037} & \textcolor{blue}{14.157820} & \textcolor{blue}{14.230763} & \textcolor{blue}{14.228753} & \textcolor{blue}{14.248626} & \textcolor{blue}{14.313685} & \textcolor{blue}{14.384037} & \textcolor{blue}{14.385648} \\ 
Re3 & \textcolor{blue}{14.041377} & \textcolor{blue}{14.085657} & \textcolor{blue}{14.096086} & \textcolor{blue}{14.147831} & \textcolor{blue}{14.209142} & \textcolor{blue}{14.270193} & \textcolor{blue}{14.269175} & \textcolor{blue}{14.335754} & \textcolor{blue}{14.397757} \\
Re4 & \textcolor{blue}{14.057577} & \textcolor{blue}{14.060922} & \textcolor{blue}{14.077708} & \textcolor{blue}{14.130088} & \textcolor{blue}{14.187400} & \textcolor{blue}{14.227497} & \textcolor{blue}{14.264671} & \textcolor{blue}{14.330714} & \textcolor{blue}{14.385614} \\
1b & 6.544452 & 6.550127 & 6.559783 & 6.562215 & 6.552776 & \textcolor{red}{6.380268} & \textcolor{red}{6.387583} & \textcolor{red}{6.390818} & \textcolor{red}{6.374640} \\
2b & 6.473904 & 6.489662 & 6.494441 & 6.478242 & 6.481587 & \textcolor{red}{6.306781} & \textcolor{red}{6.310824} & \textcolor{red}{6.293594} & \textcolor{red}{6.293976} \\
3b & 6.453137 & 6.460951 & 6.460355 & 6.466402 & 6.465974 & \textcolor{red}{6.283784} & \textcolor{red}{6.283642} & \textcolor{red}{6.283117} & \textcolor{red}{6.286072} \\
4b & 6.429984 & 6.431740 & 6.432906 & 6.435708 & 6.434777 & \textcolor{red}{6.266458} & \textcolor{red}{6.267746} & \textcolor{red}{6.262848} & \textcolor{red}{6.260575} \\
\hline
\end{tabular}}
\end{table}
\end{center}

In order to investigate the redistribution of the electron density due to S$\leftrightarrow$Se exchange, we perform Bader charge analysis \cite{bader_charge, henkelman_cms_2006}. In this approach, the volume containing the material (here, its unit cell) is divided into atomic (Bader) volumes which contain a single density maximum and are separated from other regions by surfaces on which the charge density is a minimum normal to the surface. Importantly, as Bader partitioning is based upon the charge density, it is insensitive to the basis set used.

We show in Table \ref{bader_charge} the calculated Bader charges (charges contained within each Bader volume which can be identified with a specific atomic site) for the pristine ReX$_{2}$ as well as all the most stable Janus monolayers. To identify the atomic sites, we use the numering 1-4, as in the main text, for the chalcogen sites, with the letter t or b denoting the top (above the transition metals) or bottom (below transition metals) chalcogen layer. For the pure compounds, ReS$_{2}$ and ReSe$_{2}$, inversion symmetry maps a site $i$t, $i=1,2,3,4$, onto $i$b. In turn, we number the rhenium sites as shown on the top of Fig.~\ref{fig:cohp}. For a given Janus monolayer, we indicate with black, blue and red sites occupied by sulphur, rhenium and selenium, respectively. For all compounds, the total Bader charge sums up to the total number of valence electrons in the system, 108 (15 for rheniums and 6 for sulphurs and seleniums), with the accuracy of 0.0005 of electron charge. As seen in the table exchange of S for Se at a site leads to lowering of the electron density around that site to the level close to that found at that location in ReSe$_{2}$. The electron density pushed away from a chalcogen site as a results of such substitution is mainly transferred to the rhenium sites. This means that, on an interatomic scale, charge redistribution is only weakly affected by the detailed composition of a particular monolayer.
